\documentclass{article}

\usepackage[a4paper, total={6in, 8in}]{geometry}

\usepackage{authblk}
\usepackage{url}
\usepackage{color}
\usepackage{graphicx}
\usepackage{amsmath}

\definecolor{red}{RGB}{255,0,0}

\begin{document}

\title{Inferring high-resolution human mixing patterns for disease modeling}

\author[1]{Dina Mistry}
\author[2,3]{Maria Litvinova}
\author[2]{Ana Pastore y Piontti}
\author[2]{Matteo Chinazzi}
\author[4]{Laura Fumanelli}
\author[5]{Marcelo F. C. Gomes}
\author[2]{Syed A. Haque}
\author[6]{Quan-Hui Liu}
\author[2]{Kunpeng Mu}
\author[2]{Xinyue Xiong}
\author[7,8]{M. Elizabeth Halloran}
\author[9]{Ira M. Longini Jr.}
\author[4]{Stefano Merler}
\author[4]{Marco Ajelli$^\dag$}
\author[2,3]{Alessandro Vespignani$^\dag$}

\affil[1]{Institute for Disease Modeling, Bellevue, WA, USA}
\affil[2]{Northeastern University, Boston, MA, USA}
\affil[3]{Institute for Scientific Interchange Foundation, Turin, Italy}
\affil[4]{Bruno Kessler Foundation, Trento, Italy}
\affil[5]{Funda\c{c}\~{a}o Oswaldo Cruz, Rio de Janeiro, Brazil}
\affil[6]{College of Computer Science, Sichuan University, Chengdu, Sichuan, China}
\affil[7]{Fred Hutchinson Cancer Research Center, Seattle, WA, USA}
\affil[8]{Department of Biostatistics, University of Washington, Seattle, WA, USA}
\affil[9]{Department of Biostatistics, College of Public Health and Health Professions, University of Florida, Gainesville, FL, USA}

\maketitle
$^\dag$ To whom correspondence should be addressed.\\

\begin{abstract} 
Mathematical and computational modeling approaches are increasingly used as quantitative tools in the analysis and forecasting of infectious disease epidemics. The growing need for realism in addressing complex public health questions is however calling for accurate models of the human contact patterns that govern the disease transmission processes.  Here we present a data-driven approach to generate effective descriptions of population-level contact patterns by using highly detailed macro (census) and micro (survey) data on key socio-demographic features. We produce age-stratified contact matrices for 277 sub-national administrative regions of 
countries covering approximately 3.5 billion people and reflecting the high degree of cultural and societal diversity of the focus countries. We use the derived contact matrices to model the spread of airborne infectious diseases and show that sub-national heterogeneities in human mixing patterns have a marked impact on epidemic indicators such as the reproduction number and overall attack rate of epidemics of the same etiology. The contact patterns derived here are made publicly available as a modeling tool to study the impact of socio-economic differences and demographic heterogeneities across populations on the epidemiology of infectious diseases.
\end{abstract}
Keywords: Human mixing patterns | Computational modeling | Infectious diseases | Influenza
\vskip1cm

Mathematical and computational models of infectious disease transmission are increasingly used to provide situational awareness and forecasts during epidemic outbreaks and quantitative answers to complex public health questions such as evaluating the effectiveness of control strategies (vaccination, school closure, etc.) \cite{VAN12}. Modeling approaches have thus moved away from the classic homogeneous and stylized framework \cite{AND91,MET15}, progressively incorporating heterogeneities that depend on between- and within- country population variability, disease time scale, transmission settings, as well as the specific pathogen characteristics. For instance, geographically structured models allow evaluation of spatially heterogeneous interventions in both animal and human diseases \cite{KEE03,COL07}, while individual-based models lay down the possibility of simulating all micro-details of the transmission process and tracking in time and space each individual of the simulated population \cite{LON05,FER05,MER10}. These data-driven approaches have highlighted the importance of the social, demographic, and economic characteristics of the population in determining the actual mesh of contacts underlying disease spreading among individuals. For this reason a broad range of methodologies has been used to study human mixing patterns, including surveys \cite{FEN01}, contact diaries \cite{WAL06,MOS08,HEN09,HOR11,REA14,AJE17,MEL17}, wearable sensors \cite{CAT10,KIT16}, analysis of time-use data \cite{ZAG08}, development of synthetic populations \cite{FUM12,GRE13,GAL18}, and mixed approaches \cite{IOZ10,PRE17,LIT19}, although a general modeling framework is still lacking because contact patterns among individuals vary according to the geographical scale (from census blocks to the national level), the disease under consideration, and the detailed socio-economic and demographic characteristics of the population. 

Here we present a data-driven approach to generate effective descriptions of complex contact patterns that can be used to inform infectious disease modeling approaches, including the widely adopted compartmental modeling framework. We make use of highly detailed macro (census) and micro (survey) data from publicly available sources on key socio-demographic features (e.g., age structure, household composition and members' age gaps, employment rates, school structure) to construct  synthetic populations of interacting agents, each one representing an hypothetical individual in the real population. The proposed method relies on both macro- and micro-level data for multiple socio-economic characteristics and can be adapted to different geographical contexts and diseases; something that is not possible in a ``one-model-fits-all'' approach.

We provide synthetic contact networks for nations around the world with substantially large and diverse populations. Specifically, we report contact patterns  at the subnational level in the following countries: Australia, Canada, China, India, Israel, Japan, Russia, South Africa, and the United States of America. These populations account for 277 sub-national administrative regions (such as states, provinces, prefectures, territories, etc. depending on the considered country), cover nearly 38\% of the world's surface area, and account for approximately 3.5 billion people of the world's 7.6 billion population. The resulting synthetic populations are used to derive a mesoscopic description of the human contact patterns by defining age-stratified  contact matrices for the most common social settings in which individuals spend their time interacting with each other (i.e., households, schools, workplaces, and the general community). The resulting contact matrices capture differences at the subnational level that reflect the high degree of cultural and societal diversity of the focus countries.

\begin{figure*}[t!]
\begin{center}
\centerline{\includegraphics[width=.999\textwidth]{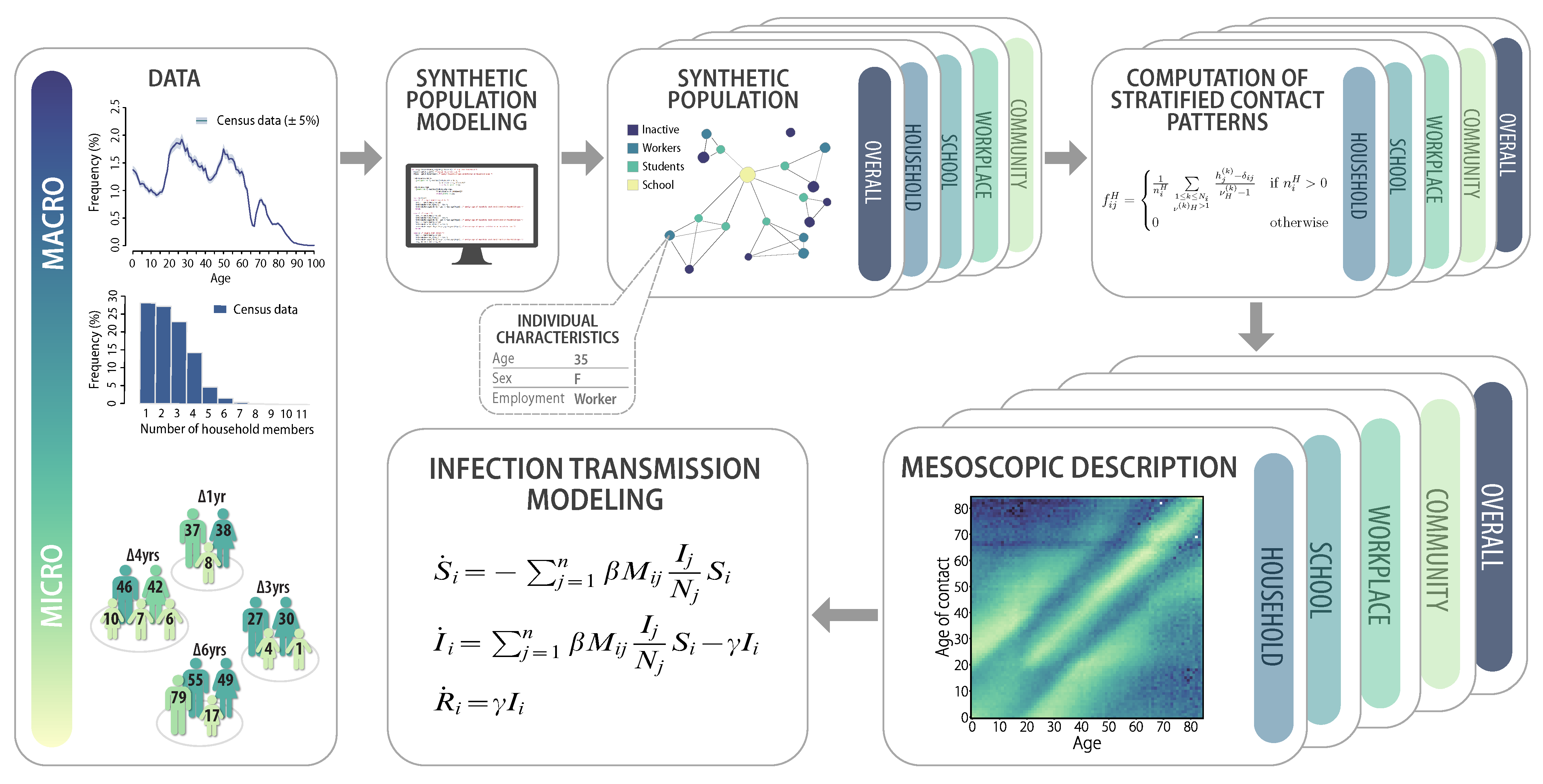}}
    \caption{{\bf Modeling framework.} Schematic representation of the workflow for modeling human mixing patterns and infection transmission dynamics.}
        \label{fig:methods}    
\end{center}     
\end{figure*}

To illustrate the importance of considering national and sub-national heterogeneities in the analysis of infectious disease epidemiology, we construct the overall contact matrix relevant for airborne infectious diseases by calibrating the combination of setting-specific contact matrices using as ground truth seven diary-based contact matrices (six European countries \cite{MOS08} and Russia \cite{AJE17}). The resulting matrices are validated against out of sample contact data collected in France \cite{BER15}, Japan \cite{MUN19}, and China \cite{ZHA19}. These contact matrices are then used in the modeling of influenza transmission patterns at national and sub-national level. The influenza modeling simulations, although considering identical disease etiology, highlights considerable heterogeneities in reproduction number and attack rates across regions of the world reflecting differences in key demographic drivers such as average age and student population. 

As a service to the community, a database containing the inferred setting-specific matrices as well as the overall contact matrices for all locations (and countries) is available on the dedicated website: https://github.com/mobs-lab/mixing-patterns. Python codes to work with the contact matrices and examples of how to use them in age-structured compartmental models are available on the same website as well. This presented work  can be easily generalized to other countries and settings, and arm the community with a general framework that can be used to make inference on important epidemiological parameters in the modeling of infectious diseases. 

\begin{figure*}[t!]
\begin{center}
\centerline{\includegraphics[width=.999\textwidth]{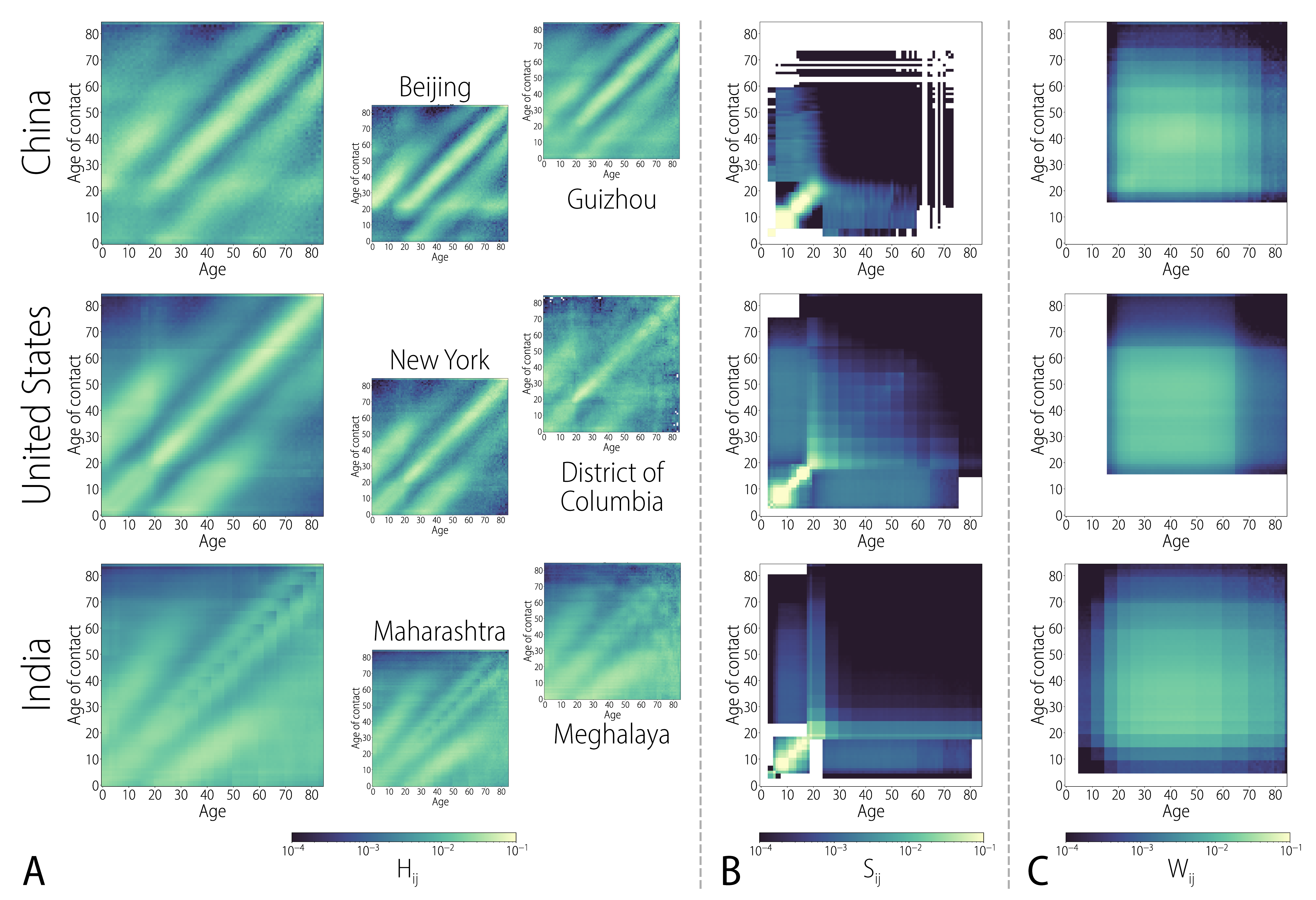}}
    \caption{{\bf Age mixing patterns by setting.}  Each heatmap represents the average frequency of contact between an individual of a given age (x-axis) and all of their possible contacts (y-axis). {\bf A}  Matrices of household contacts by age at the national level for China, the United States, and India. The six smaller panels in the center and on the left show household contact matrices at the subnational level in two provinces of China (Beijing, Guizhou), two locations of the United States (the state of New York and the District of Columbia), and two states of India (Maharashtra, Meghalaya). {\bf B}  Matrices of school contacts by age at the national level (from top to bottom: China, the United States, India). {\bf C} Matrices of work contacts by age at the national level (from top to bottom: China, the United States, India).}
        \label{fig:householdsCM}    
\end{center}     
\end{figure*}

\section*{Results} 
We use a data-driven computational approach to infer the contact networks in the social settings where people interact and spend most of their time. In particular, we focus on four social settings (household, school, workplace, and the general community), which are particularly relevant for influenza transmission \cite{FER05,AJE14}. To reconstruct the synthetic population in each context we use a wide variety of national and sub-national micro-level, census and demographic data  that provide the guidelines not only on separate characteristics of the population, but also on the interconnection of multiple characteristics. Microlevel data extrapolated from socio-demographic surveys is especially useful as no assumptions on the rules of disaggregation are required (the data is already on the required level of disaggregation). 

Contacts between individuals in the real-world populations are inferred by analysis of the generated data-driven synthetic networks by measuring the frequency of links between individuals (living, going to school, or working together) in the synthetic contact networks of the different social settings. Then we compare summary statistics derived from the generated synthetic population for each geographical area to those reported in official (macro) statistics (e.g., census data). Examples of the summary statistics used in the approach are the age structure of the population, distributions of household size, type, number of children by household size, and so on, depending on the summary statistics available from official sources. The generated data is compared to the distributions of summary statistics by using goodness of fit tests at the desired level of significance (generally 5\%). We use a non-parametric bootstrap procedure to test the uncertainty level of our sampling. This procedure is iterated until a satisfactory fit is reached. In the case of inadequate micro data (e.g., sub-optimal sample size), we use the available micro data to extrapolate rules on the age gaps between household members conditioned on the age of the household head, household size, and the relation between the members (e.g., age gap between spouses, age gap between siblings). Note that the same arguments are extended to other settings (e.g., schools, workplaces, hospitals) and can be extended to further stratifications relevant for other diseases (e.g., easy access to health care facilities). An illustration of the matrices construction workflow is reported in Fig.~\ref{fig:methods}, while the full technical description is reported in the Methods section and the Supplementary Material (SM). 

\subsection*{Setting specific contact matrices}\label{sec: CM_matrices}
We report here the results for populations of 277 sub-national administrative regions of Australia, Canada, China, India, Israel, Japan, Russia, South Africa, and the United States of America, characterizing contact patterns for about  3.5 billion individuals. We also include data at the national level for 26 European countries~\cite\cite{FUM12}. The inferred age-specific contact matrices reveal striking patterns, of which many are common to the diverse locations under study. Figure ~\ref{fig:householdsCM} shows the age mixing patterns  $F_{ij}^k$ defined as the per capita frequency of contact of an individual of age $i$ with an individual of age $j$ in setting $k$.  

Starting with the households setting in Fig.~\ref{fig:householdsCM}A, we observe that contacts between individuals can largely be characterized as that between couples living together, and parents and their children in the same household~\cite{MOS08,FUM12}. The increased frequency of contact between adults of similar ages along the main diagonal of the household contact matrix represents couples of similar ages living together, while the bands of high frequency above and below the main diagonal indicate contact between parents and children. While most locations share these overall features, the contact matrices show different age-mixing patterns. For instance, in China (Fig.~\ref{fig:householdsCM}A), the lower frequency of contact between children within households is the reflection of the country's so-called ``One-child policy''. The policy, enacted in 1979, has resulted in over a generation of many Chinese youth growing up without siblings, and hence having less contact on average with other children in this setting. This is in stark contrast with the United States and India (Fig.~\ref{fig:householdsCM}A), where the presence of multiple children born to a family results in an increased frequency of contact between this age group in the household matrix. The presence of multigenerational families in countries like India is also evident from the increased frequency of contact between all age groups, notably between the elderly (60 years and older) and young children. Even within the same country, contact patterns may be markedly different. Fig~\ref{fig:householdsCM}A shows the age-mixing patterns within households for two different provinces of China: Beijing and Guizhou. While the household contact patterns in Beijing show a clear signal of the ``One-child policy'', Guizhou shows the presence of multigenerational families, as well as  an increased presence of multiple children living in the same household. This can be traced back to the fact that the Guizhou Province is characterized by a large frequency of minority groups and the ``One-child policy'' was less strictly applied for minorities. 

Fig~\ref{fig:householdsCM}B,C show the inferred contact matrices in the school and workplace settings for China, the United States, and India. In both settings, the age mixing patterns vary strongly, reflecting differences in the educational systems, and economic conditions unique to each location. For all locations in our study, the school setting consistently exhibits the highest frequencies of contact between children and young adults attending school together. Interaction with older adults in this setting reflects the contact students have with instructors and other staff members in school. The variability of age mixing patterns between children in India (Fig~\ref{fig:householdsCM}B) also reflects the many different kinds of schools that children can attend throughout the country and the different age groups found in those schools. In the workplace environment most interaction takes place between individuals in the range of 20 to 65 years of age, with the age range depending on local retirement, employment regulations, and culture. For instance, in many parts of the world it is common for teenagers to be fully or partially employed (see the work contact matrix for the US - Fig.~\ref{fig:householdsCM}C); in India, census records for employment list even children among the population of workers. 

Statistical validation of the contact matrices against summary statistics of a large set of socio-demographic indicators has been performed to validate our results (see SM).

\begin{figure*}[t!]
\begin{center}
\centerline{\includegraphics[width=.999\textwidth]{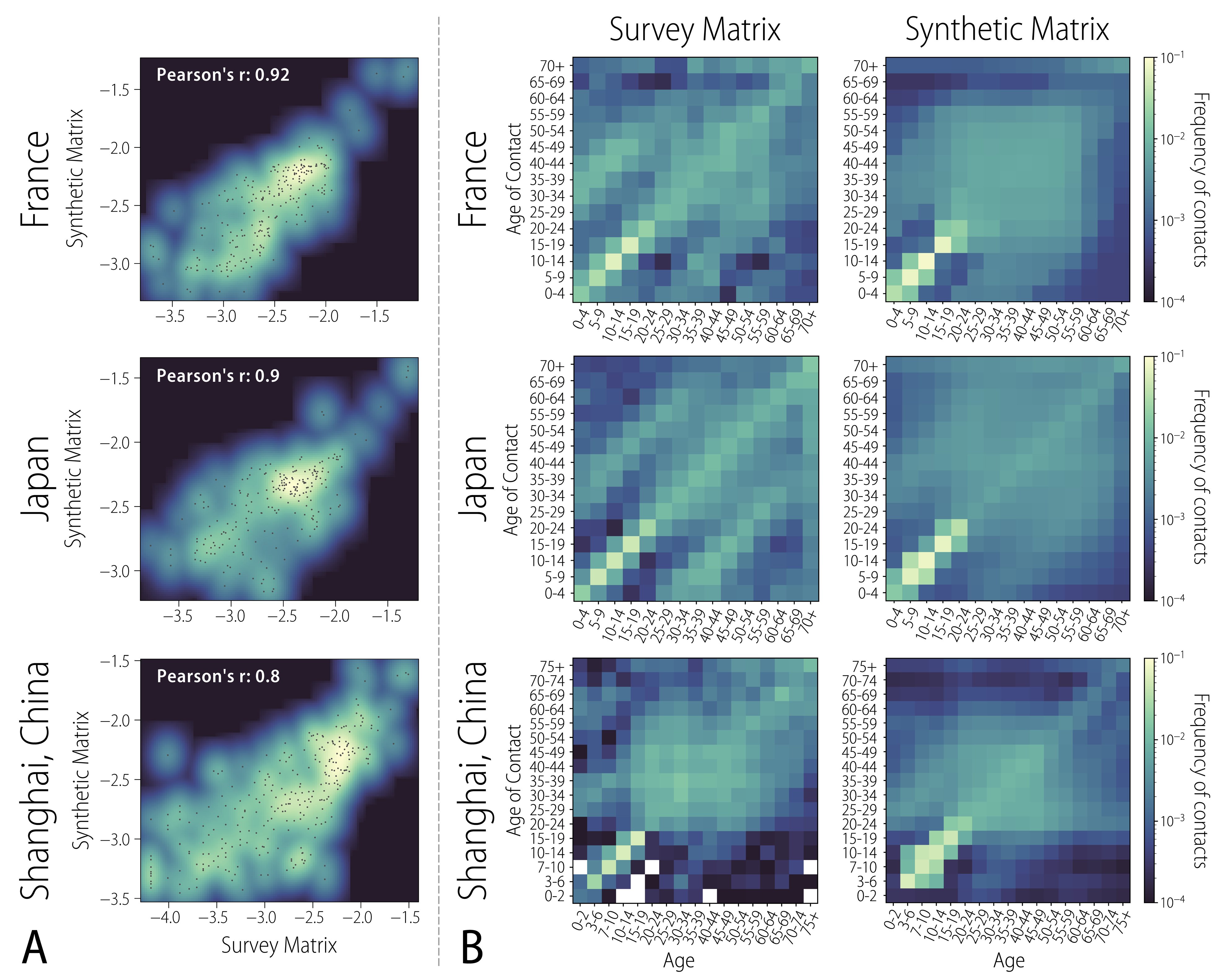}}
    \caption{{\bf Comparison to out-of-sample survey matrices.} {{\bf A} Density plots showing the correlation of survey-based contact matrices for three out of sample locations (France, Japan, and the Shanghai Province of China) and their respective synthetic contact matrices (all normalized to sum to one). The points represent the actual values of the survey and synthetic contact matrices. The linear correlation between the elements of each survey matrix and the corresponding elements of the synthetic matrix is reported in terms of Pearson correlation coefficient, whose values are reported in each plot. {\bf B} Heatmaps representing the normalized survey matrices and the normalized overall synthetic matrices for France, Japan, and the Shanghai Province of China.}}
     \label{fig:out_of_sample}
\end{center}    
\end{figure*}

\subsection*{Human mixing patterns for influenza transmission}\label{sec: CM_flumatrix}
The contact matrices obtained in each settings acquire epidemiological relevance when combined together to generate the effective descriptions of human mixing patterns relevant to the spreading of a specific disease. Here we define the matrix of effective contacts relevant to influenza transmission based on the relative contribution of the household, school, and workplace. In addition to these three social settings, we consider also the contribution of less structured causal encounters in the population \cite{AJE08}, by considering a community contact matrix that assumes individuals as potentially fully mixed \cite{FUM12}. To combine the different matrices, we propose a weighted linear combination of the derived matrices for the four considered social settings, and compute the overall matrix of contacts between individuals of age $i$ and individuals of age $j$, $M$ (whose elements are denoted as $M_{ij}$), as a weighted linear combination of setting specific contact matrices:
\begin{equation*}
	M_{ij} = \sum \limits_{k} \omega_k F_{ij}^{k}
	\label{eqn:totmatrix}
\end{equation*}
where the element $M_{ij}$ represents the average number of contacts with individuals of age $j$ for an individual of age $i$ per day, and each $\omega_k \geq 0$ is indicating the number of contacts in each setting $k$. 

Generally the $\omega_k $ are unknown disease-specific weights accounting for the relative importance of the different social settings in the transmission of a specific infectious disease. In the case of airborne infectious diseases we leverage data from diary-based survey contact matrices reported in \cite{MOS08, AJE17} for Finland, Germany, Italy, Luxembourg, The Netherlands, the United Kingdom, and the Tomsk Oblast region of Russia. Namely we perform a multiple linear regression analysis to find the values of $\omega_k$ such that the resulting  $M_{ij}$ best fits the empirical data. Note that the empirical matrices derived in \cite{MOS08,AJE17} describe the average number of contacts of age $j$ for an individual of age $i$, and in the Methods section we show how $\omega_k $ is related to  average number of contacts $<c>$ per individual. The regression yields 4.11 contacts (standard error, SE 0.41) in the household setting, 11.41 contacts (SE 0.27) in schools, 8.07 contacts (SE 0.52) in workplaces, and 2.79 contacts (SE 0.48) for the general community setting. It is worth remarking that this approach provides overall best matching $\omega_k$ and that, in principle, some of the difference of social behavior of specific countries may not be captured by this approach. For this reason, as a validation of this calibration method, in Fig.~\ref{fig:out_of_sample}A we report the correlation between the resulting synthetic matrices for France, Japan, and the Shanghai Province of China and the available empirical matrices for these additional locations \cite{BER15,MUN19,ZHA19}. We find significant ($p$-value $<0.001$) Pearson correlations of 0.92, 0.9, and 0.8 for France, Japan, and Shanghai Province, respectively. Moreover, we use the Canberra distance as a measure of the similarity between two contact matrices \cite{FUM12} (see Methods section for the definition of the Canberra distance). We estimate the distance between the seven survey-based matrices used in the calibration phase and their respective synthetic matrices to be 0.21 on average (range: 0.17-0.28). (Note that the resulting Canberra distance is normalized by the square of the number of elements of the contact matrix to account for the different number of age groups considered by the different diary-based contact surveys). When considering the three locations used as out-of-sample validation, we estimate a slightly larger average distance of 0.29 (range: 0.21-0.37), suggesting the adequacy of the employed methodology. Finally, Fig.~\ref{fig:out_of_sample}B shows a visual comparison between the synthetic and survey matrices, which highlights that the synthetic contact matrices are able to capture the specific features of each location such as contact patterns at school and the relative intensity of the main diagonals.

\begin{figure*}[t!]
\begin{center}
\centerline{\includegraphics[width=.999\textwidth]{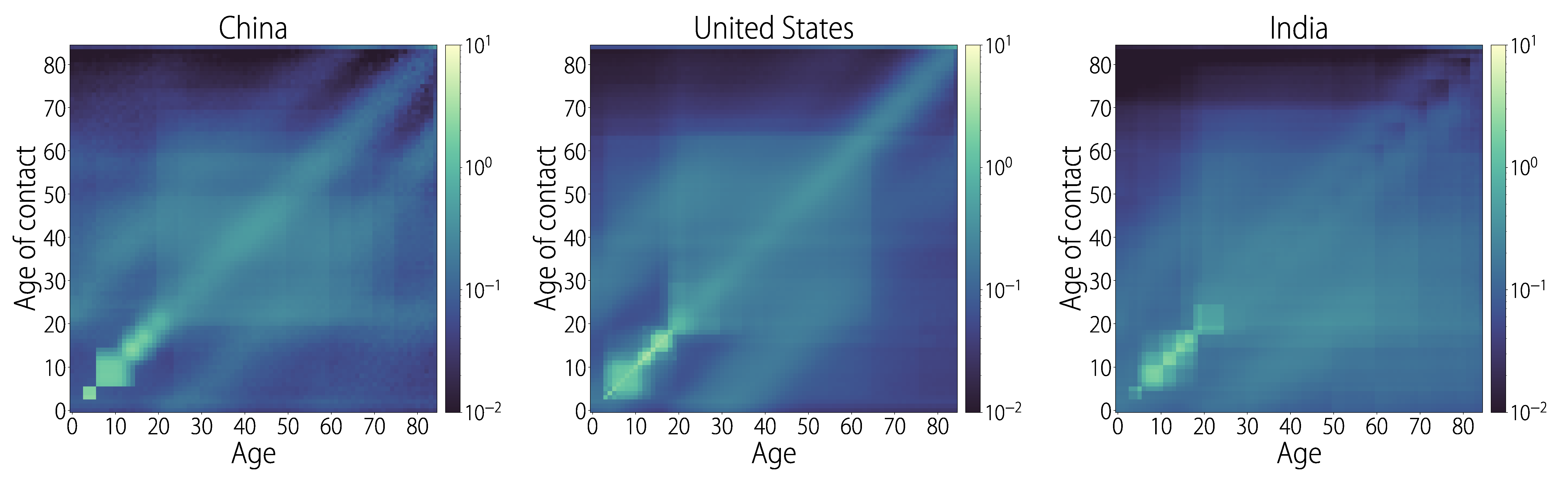}}
    \caption{{\bf Overall contact matrices.} Each heatmap represents the overall average number of contacts relevant for airborne infectious disease transmission by age at the national level for China, the United States, and India.  }
     \label{fig:fluCM}
\end{center}    
\end{figure*}

\begin{figure*}[tp!]
\begin{center}
\centerline{\includegraphics[width=.8\textwidth]{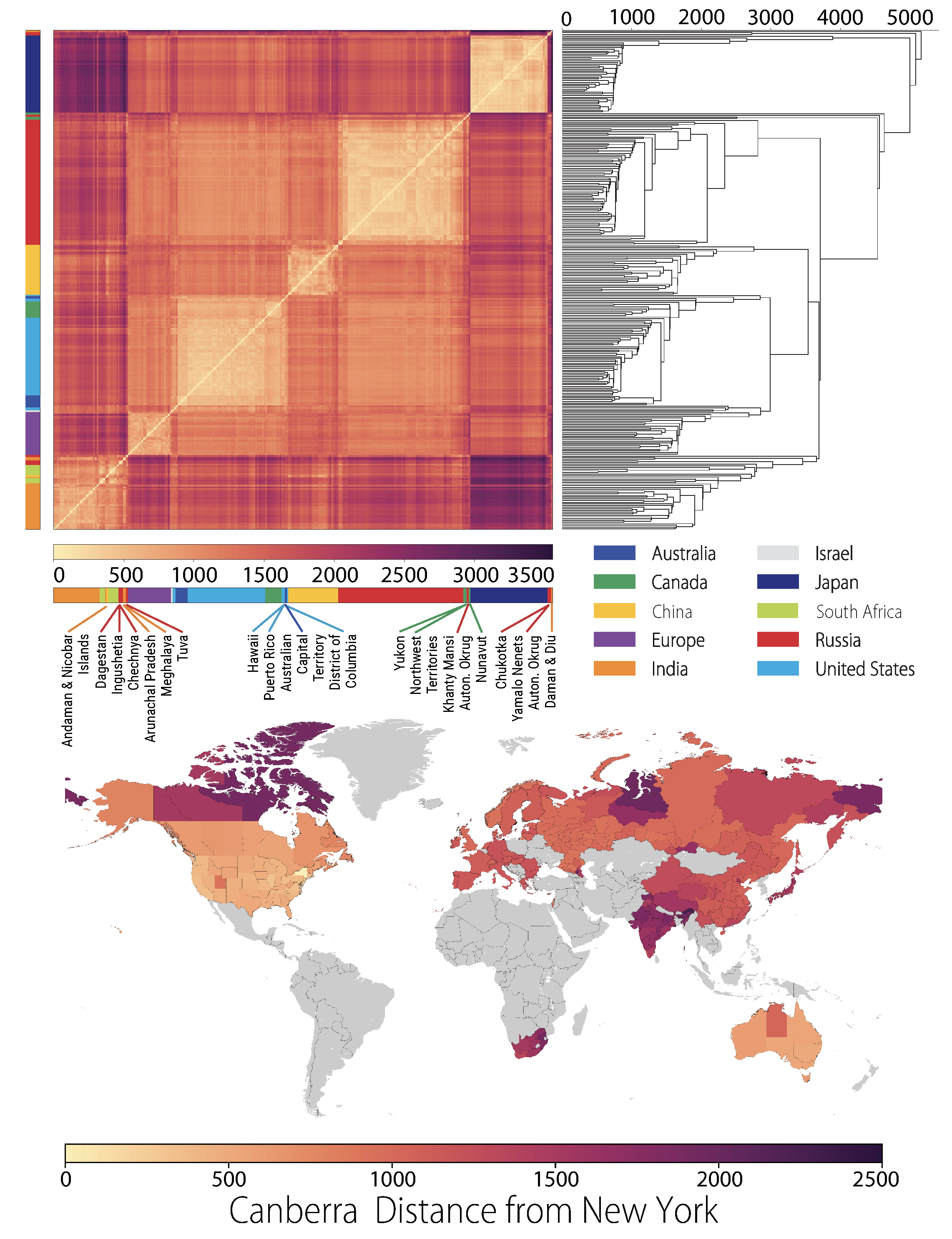}}
    \caption{{\bf Clustering of contact matrices.}  {\bf A} Clustered matrix of the Canberra distance between subnational contact matrices and associated dendrogram using hierarchical clustering to organize subnational locations. Lighter colors indicate locations more similar to each other (distance closer to 0).  {\bf B} World map of the subnational level where colors represent the Canberra distance between each subnational location and the US state of New York (used as a reference point). The gray color means that no data is available. Note that the country of Israel is treated at the national level, rather than the subnational level, due to both its relatively small population and area, and the resolution of data available for reconstruction. }
     \label{fig:clustering}
\end{center}    
\end{figure*}

Figure ~\ref{fig:fluCM}A shows the synthetic overall contact matrices for China, the United States, and India at the country level. The  contact matrices for all locations share many similarities: bands of increased contact along the main and off diagonals reflect the familiar household contact patterns, increased contact between adults age 20 to approximately 65 years old account for the interactions between the population's workforce, and the dominant contact patterns in the lower left of the contact matrices reflect the high number of interactions between school aged individuals. Depending on the age structure of the population the intensity of interactions occurring in the school setting can vary, however this feature consistently dominates the contact matrix for all locations in our study.

\begin{figure*}[t!]
\begin{center}
\centerline{\includegraphics[width=.999\textwidth]{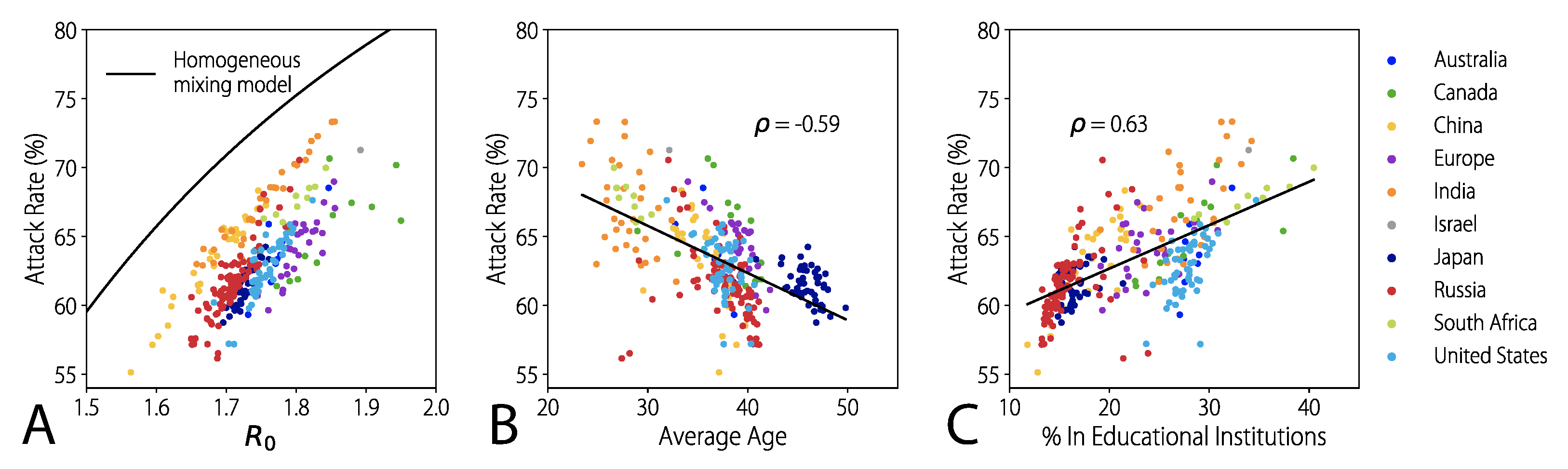}}
    \caption{{\bf Epidemic impact.} {\bf A} Scatter plot of the attack rate and the reproduction number $R_0$ from an age-structured SIR model using the contact matrix for each subnational location. European countries are included. The black line shows the results of the classic homogeneous mixing SIR model (no age groups)  {\bf B} Scatter plot of attack rates and average age in each location. The black line represents the best fitting linear model demonstrating a negative linear correlation between attack rates and average age of the population.  {\bf C} Scatter plot of attack rates and percentage of the population attending educational institutions in each location. The black line represents the best fitting linear model.}
    \label{fig:ARresults}
\end{center}    
\end{figure*}

To quantify the similarity between the overall contact matrices in different locations, we use a hierarchical clustering algorithm based on the Canberra distance to identify clusters of locations (dis)similar to each other \cite{FUM12}. We find that locations tend to cluster together by country (Fig.~\ref{fig:clustering}A), indicating that overall the contact patterns within a single country are more similar to each other than to the patterns observed in other countries. Strikingly, though not surprisingly, locations within developed countries such as Australia, Canada, and the United States are similar to each other and are clustered together, while at the same time locations throughout India, South Africa, and the North Caucasus region of Russia also cluster together, indicating a similarity in patterns between locations in the developing and transition world. Interestingly, a few territories of Canada, Russia, and India are outliers, indicating that the contact patterns in these locations are different from what is observed in all other locations (including their respective countries). A more detailed discussion is reported in the SM. If we consider the US state of New York as reference and compute the distance from all other locations to it, a geographical pattern clearly emerges (Fig.~\ref{fig:clustering}B). Indeed, the contact patterns in most states of the US, and the urbanized areas of Canada and Australia appear to be very closely related to the one inferred for New York. In contrast, most of India, South Africa, and of the territories in Canada, Russia, and Australia have contact patterns noticeably different from those obtained for the state of New York.  

\subsection*{Epidemiological relevance}\label{sec:sir_simulation}
To investigate the effect of the contact matrices on the infection transmission dynamics, we develop an age-structured SIR model to describe influenza transmission dynamics. The SIR model describes the spread of influenza in terms of the transition of individuals between different epidemiological compartments. Susceptible individuals (i.e., those at risk of acquiring the infection - S) can become infectious (i.e., capable to transmit the infection - I) after coming into contact with infectious individuals. Subsequently infectious individuals recover from the infection and become removed (R) after a certain amount of time (the infectious period). In an age structured implementation of the model, individuals are now identified also by their age, and the contact matrix is introduced to describe the number of contacts between susceptible individuals of age $i$ and all of their possible infectious contacts of age $j$ \cite{AND91,WAL06} (see Sec. Methods for details). More specifically we considered a transmission model with identical disease parameters. The contact matrices are thus the only factor driving the difference in dynamics and attack rate (total number of infected individuals) of the simulated epidemic.

\begin{figure*}[t!]
\begin{center}
     \centerline{\includegraphics[width=.98\textwidth]{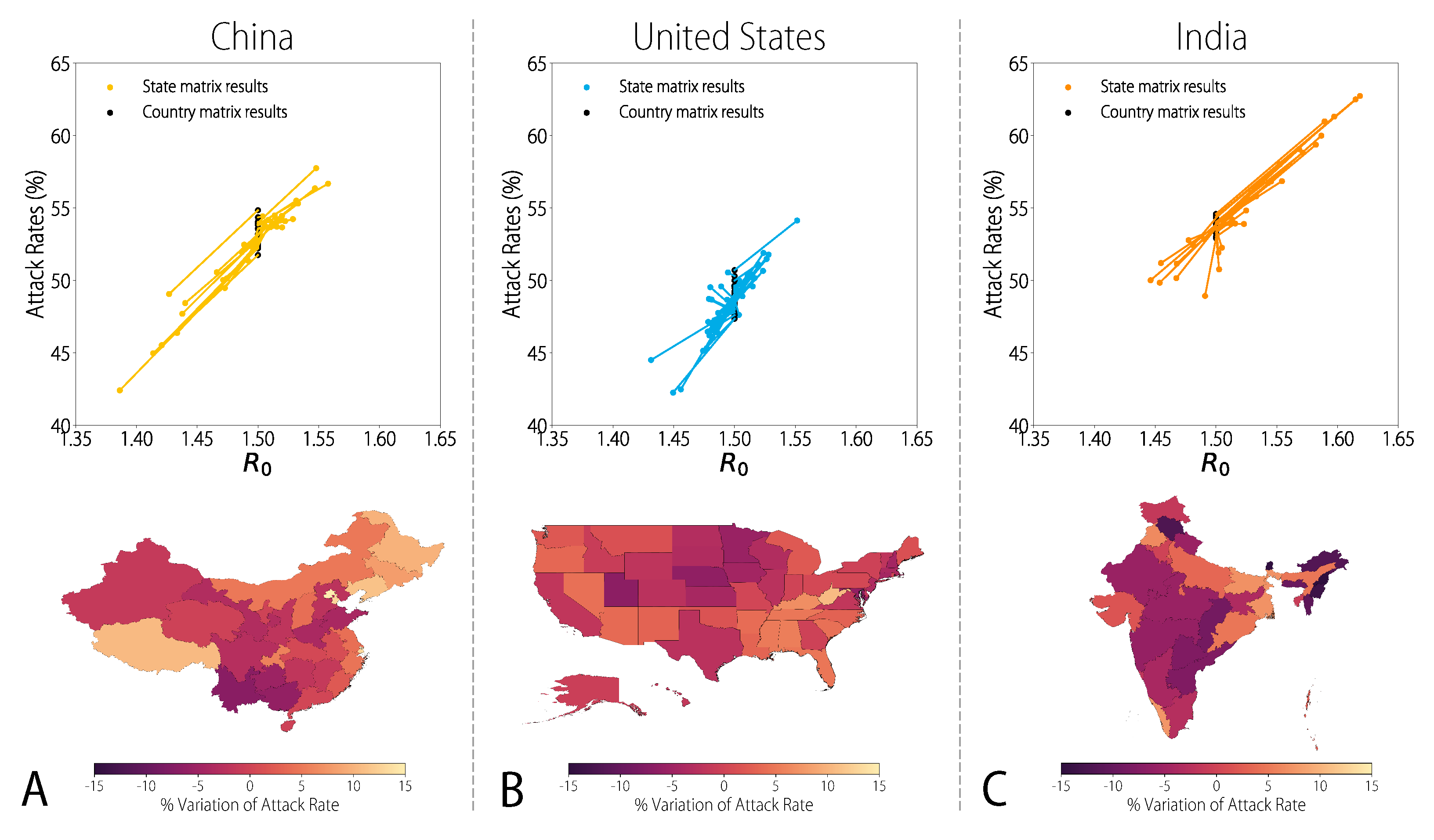}}
    \caption{{\bf Sub-national heterogeneity.} {\bf A} The black dots represent the estimated attack rates in each province of China by using the country-level contact matrix and the location-specific age structure of the population. Colored dots represent the estimated attack rates in each location by using both location-specific contact matrix and age structure of the population. The colored lines connect the two estimated values of attack rate for each location. The transmission rate is set such that $R_0=1.5$ when using the country-level matrix. Each map shows the percentage variation of the attack rate using the location-specific contact matrix with respect to using the national contact matrix as a proxy for the subnational contact patterns (i.e., $(\mbox{AR}_{c} - \mbox{AR}_{l})/\mbox{AR}_{c}$, where $\mbox{AR}_{c}$ is the attack rate estimated by using the country-level contact matrix, and $\mbox{AR}_{l}$ is that estimated by using the location-specific matrix).  Colors towards astra in the color scale indicate an overestimation of the attack rate in the location when using the country-level contact matrix as a proxy for the subnational contact patterns. Conversely, colors towards grape in the color scale indicate an underestimation of the attack rate in the location when using the country-level matrix as a proxy for the subnational contact patterns.  {\bf B} Same as A, but for the United States. {\bf C} Same as A, but for India. }
    \label{fig:country_to_states_results}
\end{center}    
\end{figure*}

Compared to the case of homogeneous mixing, where all individuals are assumed to be in contact with each other in equal proportions, the inclusion of the contact matrices in the epidemic model consistently yields a lower overall attack rate for all locations (Fig.~\ref{fig:ARresults}A). This difference is also reflected in the strong variability of the basic reproduction number $R_0$, representing the number of cases generated by a typical index case in a fully susceptible population, which depends on the spectral radius of the matrix $M$ as well as population structure (see SM). To provide further validation of the adequacy of the matrices in characterizing the specific dynamics of influenza transmission in the SM we report the simulations of the age-structured SIR model calibrated on real data from the H1N1 influenza pandemic in multiple locations. The model adequately reproduces the the age-specific seroprevalence profiles in Israel, Italy, Japan, the UK and the US \cite{WEI13,MER13,JAP10,HAR10,REE12}. 

To understand the underlying factors of the observed heterogeneities across geographical locations, we use a linear regression model to compare the attack rates and various socio-demographic features of each location (see SM). We identified two socio-demographic features that correlate strongly with the attack rate: the average age of the population (Fig.~\ref{fig:ARresults}B) and the fraction of the population in the educational system including instructors (Fig.~\ref{fig:ARresults}C). Indeed, if we examine the attack rates by age and setting (see SM), we observe that the greatest proportion of infections occur as a result of contact due to the school setting, and that attack rates in general are highest for school-aged individuals. Going further, inspection of the incidence profile by age (see SM) also clearly shows that individuals with high contact frequencies with others in the school setting are infected earlier in higher proportions. These results mirror well known influenza spreading trends/patterns observed in the real world \cite{MOS08,FUM12}. The observed results are robust (although with quantitative differences) to changes in transmissibility patterns and susceptibility to infection by age (see SM). Taken together, our results suggest that developing countries with younger populations, and thus more school-aged individuals, are likely to experience higher overall attack rates when compared to older, developed countries. 

We can also investigate how the attack rate and $R_0$ for each location would differ if we only had knowledge of the contact patterns at the national level. In this scenario, we use the country-level  influenza transmission contact matrices in each location (note that each location is still characterized by its own specific age structure) and compare the results with those obtained by using the location-specific contact matrices everything else kept identical for the disease transmission model (Fig.~\ref{fig:country_to_states_results}A-C). By using the country-level matrix we observe a much lower variability than by using location-specific mixing patterns. Moreover, location-specific attack rates and $R_0$ show a nonlinear relation with the results obtained using country-level contact patterns. Interestingly, we can observe clear geographical trends in the percent difference in attack rate using location-specific contact patterns in comparison to the corresponding country-level ones. For instance in much of the western area of China where most of the nation's ethnic minorities live, using the average matrix would lead to underestimating the final impact of an epidemic, while we would overestimate it in the more traditionally urbanized/industrialized areas in the north-east of the country, like Beijing, and Shanghai (Fig.~\ref{fig:country_to_states_results}A).

\section*{Conclusion}
We have presented a general framework for the synthetic generation of age-stratified mixing patterns in key social settings (the household, school, workplace) for the transmission of airborne infectious diseases. The contact patterns we derived are not directly measured via survey or other direct methods (e.g., wearable sensors). Rather, we infer these age-based relationships between individuals by measuring them in synthetic populations developed using a novel approach that combines macro and micro data available from public sources. While, this is a limitation as, in general, a direct measure is preferred with respect to a derived one, this approach allows us to: i) be flexible in the definition of effective contacts and thus to adapt our methodology to study of different infectious diseases which require alternative definitions of ``effective contact for transmission''; and ii) focus on broad arrays of countries for which a direct measure is not available, especially at the subnational scale.  

The use of age mixing patterns in age-structured epidemic models provides insight into the epidemiology and dynamics of infectious diseases both within and between different countries around the world, as we have shown for the case of influenza. Our approach allows the integration in modeling approaches of contact patterns that vary according to the geographical scale (from census blocks to the national level), the disease under consideration, and the detailed socio-economic and demographic characteristics of the population. The use of data-driven  heterogeneous mixing patterns, especially at the subnational level, opens up the door to potential applications in the more realistic modeling of the worldwide circulation of pathogens with epidemic/pandemic potential. The developed contact matrices also allow the study of the impact on the epidemiology of infectious diseases of socio-economic disparities and demographic peculiarities (e.g., one-child policy).  Eventually, by making all of the derived mixing patterns (in the form of readily usable contact matrices by age) publicly available, the presented results may benefit the research community actively working on the  development of infectious disease forecasting approaches and mathematical models in support of the public health decision-making processes.

\section*{Methods}
\subsection*{Development of the synthetic populations}\label{sec: CM_synpop}
To construct synthetic populations in different countries, we made use of a wide array of data sources (see SM). These data provide distributions of key socio-demographic characteristics such as the age structure, household size, age of the head of the household, age gaps between household members, household composition, employment rates, the educational system and enrollment rates, etc. Distributions such as these are typically available either as macro-level data from census databases and other governmental sources, or as micro-level data coming from surveys conducted on a sample of the population. Census databases routinely provide information at a broader scope such as the age structure of a population, or the fertility rates, however they often lack more detailed information related to the household composition and age relationships between household members. For this, we rely on micro-level surveys which collect data at the household and individual level and ask participants for information in regards to their health, household condition and composition, economic conditions, and more. The kind of data available also varies by country and even at the subnational level, thus necessitating the development of adaptive algorithms that can take in the available data and accommodate for variability in data organization to produce a faithful reconstruction of each population. With this in mind, the procedure implemented can be summarized as follows. 

The first step in the reconstruction of a real-world population is the generation of households. In this process we use two types of multinomial sampling. The first is based on the probability distribution $\mathcal{M}(y)$ of an independent socio-demographic characteristic $y$. For instance, such characteristic $y$ can be the household size or composition, depending on the data available. The second type of multinomial sampling is based on the probability $\mathcal{M}(x|y_1 = i_1,y_2=i_2,...,y_n=i_n)$ of characteristic $x$ conditional on the value $i$ of a previously determined variable(s) $y$. In this case $x$ and $y$ are assumed (when supported by available data) to have bivariate or multivariate joint distributions. Typically, the larger the number of joint distributions incorporated, the more precise the reconstruction of the real-world population. The precision of such a reconstruction is however often limited by the scope of the data (such as the survey sample size for each characteristic $y$) and its availability. For example, of the multinomial joint distributions used here one is the distribution of the age of the head of the household by the size of the household and the household composition (whether a couple, a single parent with children, siblings, multigenerational families, etc.). The bivariate joint distributions incorporated is considerably long and includes (but is not limited to) distributions of the age of household members by the age of the head of the household, the age gap between couples living together by the age of one in the pair, the mother's age at child birth by the age of the child, the number of household members by their relation to the age of the household head (such as a spouse, parent, child, grandchild, sibling, in-law, etc.) by the age of the household head and the household composition. These joint distributions were either found in the macro data or estimated from the micro survey data. Characteristics of the resulting synthetic households are compared to the distributions of the summary statistics available from the macro-level data using a goodness of fit test at the desired level of significance (generally 5\%). 

A similar procedure is used to assign those individuals to their respective schools and workplaces based on enrollment and employment records. These records detail the enrollment and employment rates by age, institutional sizes and their age structures, as well as the student-to-teacher ratios in the case of schools. A more detailed explanation of the construction of the synthetic population can be found in the SM together with the results of the comparison between the synthetic and actual population statistics.

\subsection*{Construction of age based contact matrices}\label{sec: CM_matrixeqn}
We use the synthetic contact networks to infer average age-based contact patterns within each social setting. For each location, these age-based contact patterns are encoded in a contact matrix $F^{k}$, whose elements $F_{ij}^{k}$ describes the average frequency of contact between a given individual of age $i$ and individuals of age $j$ in setting $k$. We focus on 4 social settings: the households ($H$), schools ($S$), workplaces ($W$), and the general community ($C$). Specifically, here we adopt the frequency dependent (mass action) transmission model, with the implicit assumption that an increased population density has no effect on the per capita contact rate between individuals \cite{KEE11}. This choice of modeling mechanism was already proved to represent a good approximation for the description of the transmission patterns of several infectious diseases \cite{AND91}. Moreover, it allows us to readily compare epidemiological parameters between social settings and locations with disparate population density, and thus makes for an appropriate framework when modeling the transmission dynamics of heterogeneous populations around the world. The calculation of the contact matrices can be described as follows. 

First, we compute the relative abundance of contacts between individuals of age $i$ and individuals of age $j$ in each instance $s$ of setting $k$, $\Gamma_{ij}^{k(s)}$. 
\begin{equation*}
	\Gamma_{ij}^{k(s)} = 
		\frac{\phi_i^{k(s)} (\phi_j^{k(s)} -\delta_{ij})}
		{\nu^{k(s)} - 1},
	\label{eqn:freq_matrix}
\end{equation*}
where $\phi_i^{k(s)}$ is the number of individuals of age $i$ in the instance $s$ (i.e., a specific household, school, or workplace) of setting $k$; $\delta_{ij}$ is the Kronecker delta function, which we use to omit the individual $i$ from their own set of contacts; $\nu^{k(s)}$ is the number of individuals (of all ages) in instance $s$ of setting $k$. Note that to compute $\Gamma_{ij}^{k(s)}$ we assume homogeneous mixing within each instance of the setting, and as a result the matrix $\Gamma_{ij}^{k(s)}$ has the expected symmetric property $\Gamma_{ij}^{k(s)} = \Gamma_{ji}^{k(s)}$. 

Second, we compute the per capita probability of contact of an individual of age $i$ with an individual of age $j$ in setting $k$ as $F_{ij}^k$.
\begin{equation*}
	F_{ij}^{k} = 
		\sum
		\limits_
		{\{s: \nu^{k(s)} > 1\}}
		\Gamma_{ij}^{k(s)}/N_i
	\label{eqn:avfreq_matrix}
\end{equation*}
where $N_i$ is the total number of individuals of age $i$. Note that matrix $F^{k}$ (i.e., the matrix of elements $F_{ij}^{k}$) is not symmetric. 

Third, we combine the setting-specific contact matrices by age $F^k$ to derive a matrix of the overall contacts by age $M$. We propose a weighted linear combination of the derived matrices in the four focus settings, calibrated to match the empirically estimated contact matrices from two contact diary survey studies in 7 locations throughout Western Europe and Russia \cite{MOS08,AJE17}. We perform a multiple linear regression to calibrate the weights of the synthetic setting contact matrices such that their linear combination matches the overall contact matrix for all 7 locations coming from the survey studies (see SM for details and for a comparison between the empirical and synthetic contact matrices).

\subsection*{Average number of contacts}
The average number of contacts $<c>$ can be computed as 
\begin{eqnarray*}
	<c> &=& \frac{1}{N} \sum_i N_i \sum_j M_{ij}
\end{eqnarray*}
where $N = \sum_i N_i$ is the total number of individuals in the population.
\noindent
Therefore, 
\begin{eqnarray*}
	<c> &=& \frac{1}{N}\sum_i N_i \sum_j \sum_k \omega_k F_{ij}^{k} \\
	       &=& \frac{1}{N}\sum_i N_i \sum_j \sum_k \omega_k \sum_s \Gamma_{ij}^{k(s)}/N_i \\
	       &=& \frac{1}{N}\sum_k \omega_k \sum_i \sum_j \sum_s N_i \Gamma_{ij}^{k(s)}/N_i \\
	       &=& \frac{1}{N}\sum_k \omega_k \sum_i \sum_j \sum_s \Gamma_{ij}^{k(s)} \\
	       &=& \sum_k \omega_k \sum_s \sum_i \sum_j \Gamma_{ij}^{k(s)}/N \\
	       &=& \sum_k \omega_k Z_k/N \\
\end{eqnarray*}
where $Z_k$ is the number of individuals having at least one contact in setting $k$.

\subsection*{Canberra distance} 
To make side-by-side comparisons of the inferred contact matrices by age, we use the Canberra distance \cite{FUM12}. Specifically, each matrix is treated as a vector on which the Canberra distance is defined as  

\begin{eqnarray*}
	d(x,y)_{Canberra} &=& 
	\sum\limits_{i} 
	\begin{cases} 
		\frac{|x_i - y_i|}{|x_i| + |y_i|}  
		&$ for $ x_i,y_i \neq 0\\
		1 &$ for $ x_i, y_i = 0
	\end{cases}
\label{eqn:canberra}
\end{eqnarray*}
This yields a distance value of 0 for two locations with identical contact matrices, and increasingly larger distance values for two locations with increasingly different contact matrices. 

\subsection*{Age structured disease transmission model}\label{sec: age_sir_methods}
For each location $l$ the transmission dynamics of influenza are modeled through an age-structured SIR model, where the mixing patterns are defined by the contact matrix previously introduced, $M_{ij}$. 

The model is defined by the following set of equations:  
\begin{eqnarray*}
	\label{eqn:sireqns}
		\dot{S_i} &=& -\lambda_i S_i \\
		\dot{I_i} &=& \lambda_i S_i - \gamma I_i \\
		\dot{R_i} &=& \gamma I_i
\end{eqnarray*}
where $S_i$ is the number of susceptible individuals of age $i$, $I_i$ is the number of infected individuals of age $i$, $R_i$ is the number of recovered or removed individuals of age $i$; $\gamma^{-1}$ is the infectious periods (which corresponds to the generation time in the simple SIR model \cite{WAL06b,LIU18}), which is set to 2.6 days \cite{VIN14}; and $\lambda_i$ represents the force of infection to which an individual of age $i$ is exposed to other infected individuals and expressed as
\begin{eqnarray*}
    \label{eqn:forceofinfection}
    	\lambda_i &=& \beta \sum\limits_{j} M_{ij} \frac{I_j}{N_j}
\end{eqnarray*}
where $\beta$ is the transmissibility of the infection, $N_i$ is the total number of individuals of age $i$, and $M_{ij}$ measures the average number of contacts for an individual of age $i$ with all of their contacts of age $j$. 

The basic reproduction number $R_0$, representing number of cases generated by a typical index case in a fully susceptible population, can be defined for this model as 
\begin{eqnarray*}
	\label{eqn:R0}
		R_0 &=& \frac{\beta}{\gamma} \rho(M)
\end{eqnarray*}
where $\rho (M)$ is the dominant eigenvalue of the matrix $M$ \cite{DIE90}. 

\subsection*{Data Availability}\label{sec: CM_data_availability}
A database containing the inferred setting-specific matrices as well as the contact matrices for influenza transmission for all locations (and countries) is publicly available on the dedicated website: https://github.com/mobs-lab/mixing-patterns. Python routines to work with the contact matrices and examples of how to use them in age-structured compartmental models are also available.

\section*{Acknowledgements}
APyP, MC, KM, XX, MEH, IML, and AV acknowledge funding from Models of Infectious Disease Agent Study, National Institute of General Medical Sciences Grant U54GM111274.

\addcontentsline{toc}{section}{Bibliography}

\end{document}